# Smith-Purcell Radiation

D.N. Klochkov[1], M.A. Kutlan[2*]

[1] Prokhorov General Physics Institute, Moscow, Russia

[2] Institute for Particle & Nuclear Physics, Budapest, Hungary

[*] kutlanma@gmail.com

**Abstract:** The simplest model of the magnetized infinitely thin electron beam is considered. The basic equations that describe the periodic solutions for a self-consistent system of a couple of Maxwell equations and equations for the medium are obtained.

## 1. Overview

There is currently substantial interest in the development of THz sources for applications to biophysics, medical and industrial imaging, nanostructures, and materials science [1]. At the present time, available THz sources fall into three categories: gas lasers, solid-state devices, and electron-beam driven devices. Optically and electrically pumped molecular gas lasers provide hundreds of lines between 40 and 1000 $\mu$m, but they are inherently not tunable. Solid-state THz sources include *p*-type germanium lasers, quantum-cascade lasers, and excitation of numerous materials with subpicosecond optical laser pulses. *p*-type Ge lasers can be continuously tunable from 1 to 4 THz, but require a large (1-T) external magnetic field, and must be operated at 20 K [2]. Electron-beam driven sources of THz radiation include backward-wave oscillators (BWOs), synchrotrons, and free-electron lasers (FELs). BWOs are commercially available, compact devices that operate from 30–1000 GHz and produce milliwatts of average power [3,4]. Modern synchrotrons with short electron bunches, such as BESSY II in Berlin [5] and the recirculating linac at Jefferson Laboratory [6], produce broadband radiation out to about 1 THz with tens of watts of average power. Usual FELs are operating too in several laboratories in THz region with average power more than 100 W [7-10]. Coherently enhanced THz emission from relativistic electrons in an undulator has been observed at ENEA-Frascati [11]. However, synchrotrons and conventional FELs require large facilities. There are numerous publications devoted to FELs on undulators and strophotrons [12-49 and references therein].

Another source of radiation in the THz region is Smith-Purcell (SP) radiation. The effect discovered in 1953 by Smith and Purcell [50] and called then by their names consists in emission of light by electrons moving above a grating. Later, after the advent of Free Electron Lasers (FEL), it was suggested to use the Smith-Purcell effect to create a new type of FEL – the Smith-Purcell Free Electron Laser (SP FEL).

The spontaneous Smith-Purcell (SP) radiation is generated when an electron or another charged particle passes close to the surface of a periodic structure of some conducting material, i.e., a diffraction grating. The radiation mechanism was predicted by Frank in 1942 [51] and observed in the visible spectral range for the first time by Smith and Purcell [50] in 1953 and was called SP radiation. Although the SP radiation has been studied by both experimental and theoretical methods for over 50 years since its discovery, the experimental study of it using relativistic beams was carried out only after 1992 [52-54].

An interesting opportunity for a convenient, tunable, narrow-band source is presented by the recent development of a tabletop Smith-Purcell free-electron laser (FEL) at Dartmouth [55]. This device has demonstrated super-radiant emission in the spectral region from 300-900 $\mu$ m, but barely exceeded threshold. To improve this performance, it will be necessary to develop electron beams with improved brightness and a better understanding of how these devices operate. Therefore, the experiment at Dartmouth College [55], where coherent radiation in the SP system was observed, has stimulated new investigations concerning the SP FEL as an open slow-wave structure [56-65].

In the present paper we restrict ourselves to the study of only induced Smith-Purcell instability, when the initially unbunched electron beam interacts with a wave the frequency of which is that of the SP radiation. One can select two theoretical target settings for SP radiation: the first is the problem of generation of SP-radiation. In this case no waves are incoming from the infinity. The SP system generates outgoing waves. The second target setting is that the SP-system is amplifying or attenuating the waves incoming from the infinity. One can expect that these different problems have different solutions. In the present work we consider only the problem of generation of waves by the SP-system with no incoming waves. In order that mathematical problems do not cover physical picture, we consider the simplest model of the beam, in which the infinitely thin sheet of beam stabilized by external strong magnetic field is described by hydrodynamical equations.

In this article we present the basic equations that describe the periodic solutions for a self-consistent system of a couple of Maxwell equations and equations for the medium.

## 2. Basic equations

In this section, we consider the solutions of a self-consistent system of Maxwell's and beam's equations, which are linear in the field amplitudes. We take a Cartesian coordinate system in such a way that the electron beam propagates to the positive direction of the x axis in the vacuum over the grating (parallel to the surface of the grating) as shown in Fig.1. For the sake of simplicity, we consider a sheet of the beam which is infinite in the z-direction and is stabilized by external strong magnetic field. We assume that the system has the symmetry of translational invariance in the z direction. This means that all physical values depend only on coordinates x and y, and $E_z=0$.

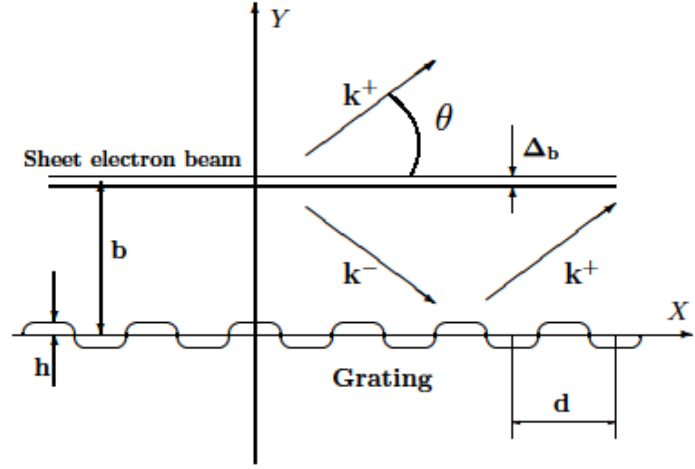

Figure 1. Schematic of a SP-FEL using a sheet electron beam. The sheet electron beam is in the plane $y=b$. $\theta$ is the angle of observation, $\Delta_b$ is the thickness of the beam, $d$ and $h$ are the period and the amplitude of the grating, respectively.

We start from Maxwell's equations:

$$\nabla \times \mathbf{E} = -\frac{1}{c}\frac{\partial \mathbf{B}}{\partial t}, \quad \nabla \mathbf{B} = 0,$$

$$\nabla \times \mathbf{B} = \frac{1}{c}\frac{\partial \mathbf{E}}{\partial t} + \frac{4\pi}{c}\mathbf{j}, \quad \nabla \mathbf{E} = 4\pi\rho \tag{1}$$

where $\mathbf{j}$ and $\rho$ are the current and charge density of the beam, respectively. We assume the simplest model of the beam, so that the initial distribution function of the electrons is expressed as

$$f_0(t,\mathbf{r}_0,\mathbf{p}_0) = \Delta_b n_{b0}\delta(y-b)\delta(\mathbf{p}_0 - m\gamma_0 \mathbf{u}) \tag{2}$$

where $\mathbf{u}=(u,0,0)$ is the initial velocity of electrons, $\gamma_0 = (1-u^2/c^2)^{1/2}$ is a relativistic factor, $\Delta_b$ is the thickness of the beam and $n_{b0}$ is its undisturbed density. For a monoenergetic electron beam we can use the hydrodynamical approach, namely, the equation of motion:

$$\frac{d\mathbf{p}}{dt} = \left(\frac{\partial}{\partial t} + \mathbf{v}\nabla\right)\mathbf{p} = e\mathbf{E} + e\frac{\mathbf{v}}{c}\times\mathbf{B} \tag{3}$$

and the continuity equation:

$$\frac{\partial \rho}{\partial t} + \mathrm{div}\mathbf{j} = 0 \tag{4}$$

where $\mathbf{j} = \mathbf{v}\rho = en_b\mathbf{v}$, $n_b$ is the beam density.

To describe the electromagnetic field in the vacuum space ($\rho = 0$ and $\mathbf{j} = 0$ in Eqs. (1), we take advantage of Floquet's theorem and expand the $\mathbf{E}$ and $\mathbf{B}$ fields as:

$$\mathbf{E}(y > b) = \sum_n \mathbf{E}_n e^{-i\omega t + i\mathbf{q}_n^+ \mathbf{r}},$$
$$\mathbf{B}(y > b) = \sum_n \mathbf{B}_n e^{-i\omega t + i\mathbf{q}_n^+ \mathbf{r}} \tag{5}$$

above the beam level, and

$$\mathbf{E}(y < b) = e^{-i\omega t} \sum_n \left( \mathbf{E}_n^+ e^{i\mathbf{q}_n^+ \mathbf{r}} + \mathbf{E}_n^- e^{i\mathbf{q}_n^- \mathbf{r}} \right),$$
$$\mathbf{B}(y < b) = e^{-i\omega t} \sum_n \left( \mathbf{B}_n^+ e^{i\mathbf{q}_n^+ \mathbf{r}} + \mathbf{B}_n^- e^{i\mathbf{q}_n^- \mathbf{r}} \right) \tag{6}$$

below the beam level. Here the wave vectors are

$$\mathbf{q}_n^\pm = (q_{nx}, \pm q_{ny}) = (k_x + n\chi, \pm q_{ny}) \tag{7}$$

Where $\chi = 2\pi/d$ is the wave number of the grating, and $d$ is its period. The sums in Eq. (5) contain only the waves outgoing from the system, while the sums in Eq. (6) contain the waves propagating in both positive and negative directions of the $Y$ axis. Since the dispersion relation for electromagnetic wave in vacuum space is $\omega^2 = \left(\mathbf{q}_n^\pm\right)^2 c^2$, we can write

$$q_{ny} = \sqrt{\frac{\omega^2}{c^2} - q_{nx}^2} = \sqrt{\frac{\omega^2}{c^2} - (k_x + n\chi)^2}. \tag{8}$$

The solution of Maxwell equations in the form of Eqs. (5) and (6) is complete, since the sums present both propagating electromagnetic and surface waves. For propagating electromagnetic waves, we find that $\mathbf{q}_0^\pm = \mathbf{k}^\pm$ $, where $\mathbf{k}^\pm = \left(k_x, \pm k_y\right)$. All waves with a plus sign propagate to the positive direction of $Y$ axis, while waves with a minus sign propagate in the direction of the negative value on the $Y$ axis. It means that the non-propagating surface waves are damped modes in the direction of their propagation, i.e. $\mathrm{Im}\, q_{ny} > 0$. The solution (5)-(8) describes the case when the waves incoming from infinity are absent, and, therefore, the solution (5)-(8) satisfies the radiated boundary condition named as the Sommerfeld boundary condition.

There are only three non-zero components of electromagnetic field $E_x$, $E_y$ and $B_z$. From the Maxwell equations for vacuum space we find

$$\left(\mathbf{E}_n \mathbf{q}_n^+\right) = 0, \mathbf{B}_n = \frac{c\mathbf{q}_n^+}{\omega} \times \mathbf{E}_n,$$
$$\left(\mathbf{E}_n^\pm \mathbf{q}_n^\pm\right) = 0, \mathbf{B}_n^\pm = \frac{c\mathbf{q}_n^\pm}{\omega} \times \mathbf{E}_n^\pm.$$
(9)

Using these relations we can express the partial amplitudes of two components in terms of the partial amplitudes of the third component. For example, we can write the partial amplitudes of $E_y$ and $B_z$ in terms of the partial amplitudes of $E_x$: $E_{ny} = -(q_{nx}/q_{ny})E_{nx}$, $B_{nz} = -(\omega/q_{ny}c)E_{nx}$, $E_{ny}^\pm = \mp(q_{nx}/q_{ny})E_{nx}^\pm$, $B_{nz}^\pm = \mp(\omega/q_{ny}c)E_{nx}^\pm$. Since this operation breaks the symmetry between the field components and will cause problems later during analytical calculations, we will take a different approach and introduce the potential $P$:

$$P = \begin{cases} \sum_n P_n e^{-i\omega t + \mathbf{q}_n^+ \mathbf{r}}, & y > b; \\ \sum_n \left(P_n^+ e^{i\mathbf{q}_n^+ \mathbf{r}} + P_n^- e^{i\mathbf{q}_n^- \mathbf{r}}\right) e^{-i\omega t}, & y < b. \end{cases}$$
(10)

Equations (4.2.10) hold if the field components are determined by relations:

$$E_x = \frac{\partial P}{\partial y}, E_y = -\frac{\partial P}{\partial x}, B_z = \frac{1}{c}\frac{\partial P}{\partial t}.$$
(11)

Equation (11) determines fields in the regions $y > b$ and $y < b$ separately. To connect the partial amplitudes $P_n$ with $P_n^+$ and $P_n^-$ we have to use the boundary condition on the beam sheet. For the beam sheet we can write:

$$\delta \mathbf{j} = \delta \mathbf{j}_s \delta(y-b) = (\delta j_{sx}, 0, 0)\delta(y-b),$$
$$\delta \rho = \delta \rho_s \delta(y-b),$$
(12)

where $\delta \rho_s$ and $\delta \mathbf{j}_s$ are the surface densities of the disturbed charge and the disturbed current, respectively. Integrating the Maxwell's equations with $\delta \rho$ and $\delta \mathbf{j}$ given by Eq. (12) over $y$ in the infinite vicinity of point $y = b$ we find the boundary conditions on the sheet beam:

$$E_x(b+0) = E_x(b-0),$$
$$E_x(b+0) - E_x(b-0) = 4\pi \delta \rho_s,$$
$$B_z(b+0) - B_z(b-0) = \frac{4\pi}{c} \delta j_s.$$
(13)

We assume that current and charge densities can be written as

$$\delta j_s = \sum_n \delta j_{ns} e^{-i\omega t + i\mathbf{q}_n^+ \mathbf{r}},$$
$$\delta \rho_s = \sum_n \delta \rho_{ns} e^{-i\omega t + i\mathbf{q}_n^+ \mathbf{r}}.$$
(14)

Then we can derive relations for the amplitudes $P_n, P_n^+$ and $P_n^-$:

$$P_n^+ e^{iq_{ny}b} - P_n^- e^{-iq_{ny}b} = P_n e^{iq_{ny}b},$$
$$P_n^+ e^{iq_{ny}b} - P_n^- e^{-iq_{ny}b} = P_n e^{iq_{ny}b} - 4\pi i \frac{\delta \rho_{ns}}{q_{nx}} e^{iq_{ny}b},$$
$$P_n^+ e^{iq_{ny}b} - P_n^- e^{-iq_{ny}b} = P_n e^{iq_{ny}b} - \frac{4\pi i}{\omega} e^{iq_{ny}b}.$$
(15)

The second and the third equations are the same if the continuity equation for the Fourier components holds: $\omega \delta \rho_{ns} = q_{nx} \delta j_{ns}$. Solutions of Eqs. (15) are

$$P_n^+ = P_n - 2\pi i \frac{\delta \rho_{ns}}{q_{nx}},$$
$$P_n^- = -2\pi i \frac{\delta \rho_{ns}}{q_{nx}} e^{2iq_{ny}b}.$$
(16)

In order to find the density $\delta \rho_{ns}$ we linearize Eqs. (3) and (4) using small perturbations. We assume that $\mathbf{v} = \mathbf{u} + \delta \mathbf{v}$, where $\delta \mathbf{v} = (\delta v_x, 0, 0)$ and $\rho = \rho_0 + \delta \rho$ with $\rho_0 = e\Delta_b n_{b0} \delta(y-b)$. The linearized equation of motion

$$\left(\frac{\partial}{\partial t} + \mathbf{u}\nabla\right)\delta v_x = \frac{eE_x}{m\gamma_0^3}$$
(17)

has solution

$$\delta v_x = \sum_n \delta v_{nx} e^{-i\omega t + i\mathbf{q}_n^+ \mathbf{r}},$$
(18)

Where

$$\delta v_{mx} = \frac{ie}{m\gamma_0^3} \frac{E_{nx}}{\omega - (\mathbf{q}_n^+ \mathbf{u})}.$$
(19)

The linearized continuity equation

$$\left(\frac{\partial}{\partial t} + \mathbf{u}\nabla\right)\delta \rho = -\rho_0 \operatorname{div} \delta \mathbf{v}$$
(20)

gives a solution for the Fourier components $\delta\rho_n$:

$$\delta\rho_n = \rho_0 \frac{q_{nx}\delta v_{nx}}{\omega - (\mathbf{q}_n^+ \mathbf{u})}. \tag{21}$$

Then we find

$$\delta\rho_{ns} = -\frac{\Delta_b \omega_b^2}{4\pi\gamma_0^3} \frac{q_{nx} q_{ny} P_n}{(\omega - (\mathbf{q}_n^+ \mathbf{u}))^2}, \tag{22}$$

where $\omega_b = (4\pi e^2 n_{b0}/m)^{1/2}$ is the Langmuir beam frequency.

The partial field amplitudes determined by Eq. (16) are

$$\begin{aligned}P_n^+ &= [1+K(\mathrm{n})]P_n, \\ P_n^- &= K(\mathrm{n})e^{2iq_{ny}b}P_n.\end{aligned} \tag{23}$$

Here we introduced the following notation

$$K(n) = \frac{i}{2}\Delta_b \omega_b^2 \gamma_0^{-3} \frac{q_{ny}}{(\omega - (\mathbf{q}_n^+ \mathbf{u}))^2}. \tag{24}$$

### 3. Conclusion

To describe the electromagnetic field in the vacuum space, we take advantage of Floquet's theorem and expand the electric and magnetic fields of electromagnetic wave in series. The coefficients of these expansions are found..

### Acknowledgments

The authors thank KB Oganesyan for helpful discussions.